\documentclass[twocolumn,showpacs,amsmath,amssymb,pre,aps]{revtex4}   
\usepackage{graphicx}
\usepackage{dcolumn}
\usepackage{bm}
\usepackage{color}

\renewcommand{\vec}[1]{\mathbf{#1}}

\newif\ifgraph

\graphtrue             %

\begin{document}
\title{Driven transport of active particles through arrays of symmetric obstacles}

\author{Shubhadip Nayak$^{1}$, Sohom Das$^{2}$,  Poulami Bag$^{1}$,  Tanwi Debnath$^{3}$ and Pulak K. Ghosh$^{1}$\footnote[2]{Email: pulak.chem@presiuniv.ac.in}}

\affiliation{$^{1}$ Department of Chemistry, Presidency University, Kolkata - 700073, India}

\affiliation{$^{2}$ Jawaharlal Nehru Centre for Advanced Scientific Research, Jakkur, Bangalore-560 064, Karnataka, India}

\affiliation{$^{3}$ Theoretical Physics of Living Matter, Institute of Biological Information Processing and \\ Institute for Advanced Simulation, Forschungszentrum J\"{u}lich, 52425 J\"{u}lich, Germany}


\date{\today}

\begin{abstract}
  We numerically  examine the driven transport of an overdamped self-propelled particle through a two-dimensional array of circular obstacles. A detailed analysis of transport quantifiers (mobility and diffusivity) has been performed for two types of channels, {\it channel I} and {\it channel II}, that respectively correspond to the parallel and diagonal drives with respect to the array axis. Our simulation results show that the signatures of pinning actions and depinning processes in the array of obstacles are manifested through excess diffusion peaks or sudden drops in diffusivity, and abrupt jumps in mobility with varying amplitude of the drive. The underlying depinning mechanisms and the associated threshold driving strength largely depend on the persistent length of self-propulsion. For low driving strength, both diffusivity and mobility are noticeably suppressed by the array of obstacles, irrespective of the self-propulsion parameters and direction of the drive. When self-propulsion length  is larger than a channel compartment size, transport quantifiers are insensitive to the rotational relaxation time.  
  Transport with diagonal drives features self-propulsion-dependent negative differential mobility. The amplitude of the negative differential mobility of an active particle is much larger than that of a passive one.
  The present analysis aims at understanding
the driven transport of active species like, bacteria, virus, Janus Particle etc. in porous medium. 
\end{abstract}
 \pacs{82.70.Dd 
87.15.hj 
05.40.Jc} \maketitle

\begin{figure}
\centering
\includegraphics[width=0.47\textwidth,height=0.25\textwidth]{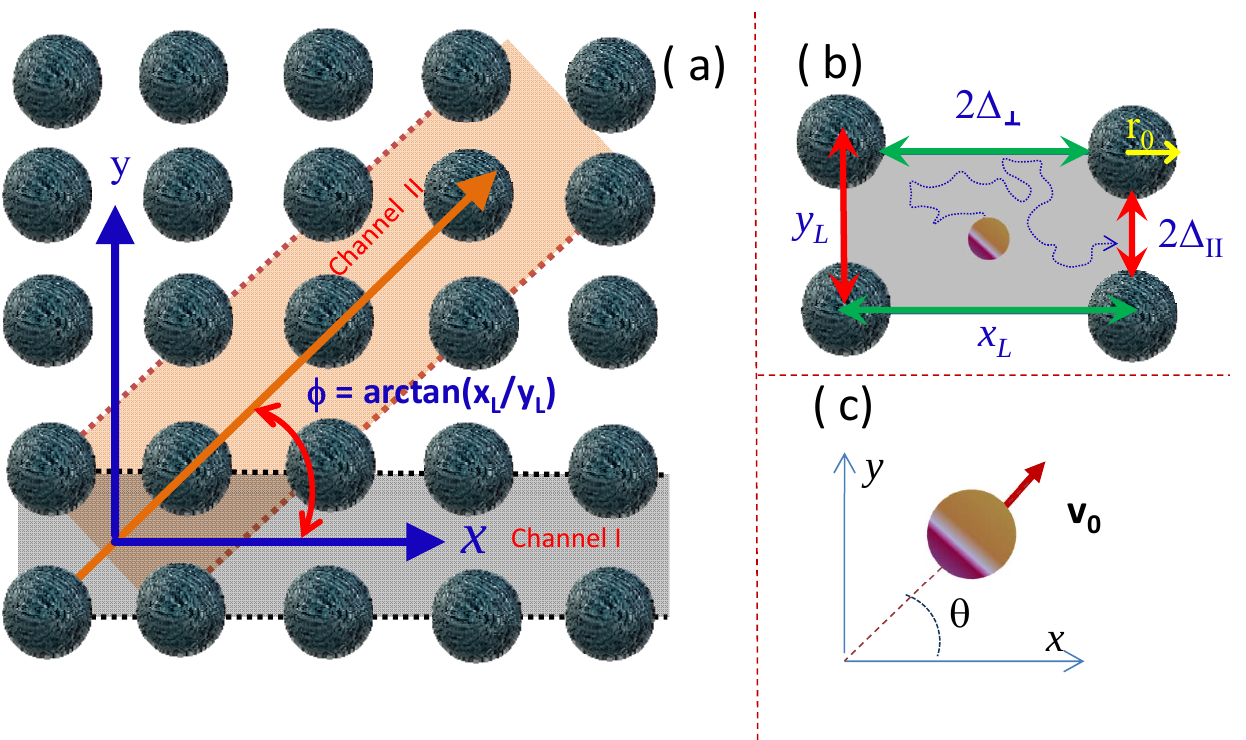}
\caption {(Color online) (a) Schematic of symmetric square array of  obstacles with unit cell $x_L \times y_L$ and obstacle radius $r_0$. Transport with a drive along  horizontal (diagonal) to array axis equivalent to transport along channel I (channel II). (b) A sketch of a compartment in channel I depicting width $y_L$, period $x_L$ and bottle necks $2\Delta_\parallel = y_L-2r_0, \;  2\Delta_\perp = x_L-2r_0 $.   (c) Sketch of an active particle of Janus kind showing direction of the instantaneous self-propulsion velocity ($\vec{v}_0$) with respect to the array x-axis.}
\end{figure}

\section{Introduction} 
Transport control of active particles through micro- and nano-structures requires detailed knowledge of diffusion mechanisms under non-holonomic conditions, accounting for uneven boundary effects. Such situations often arise in manipulating artificial active particles as a Brownian tracer  (e.g., targeted drug delivery~\cite{soto,zhou}, robotic microsurgery~\cite{ACS-Appl}, monitoring dynamics of sluggish particles through motility transfer~\cite{Engay,Debnath}), intra- and inter-cellular transport of active species in biological systems~\cite{Micali,Miller}, diffusion in porous materials, carrier diffusions in artificial photo-synthetic structures~\cite{photo-1,photo-2,photo-3,photo-4}, and sorting active particles taking advantage of uneven boundary effects~\cite{Guo-hao,JPCL}. Transport control of artificial swimmers in porous media is an crucial issue in emerging medical and nanotechnological applications.

Artificial active particles (also known as micro/nano-swimmers, micro-rockets, biobots, etc.) are often considered as synthetic analog of natural microswimmers. They are cleverly designed colloidal Brownian particles capable of self-propulsion through some self-phoretic process~\cite{review}. The simplest class of such entities are so-called Janus particles (JP) with two differently fabricated hemispheres, or ``faces". Such two-faced particles can create either a concentration gradient by assisting some chemical processes on the surface of the active hemisphere~\cite{cataly1,cataly2,cataly3}, or a thermal gradient~\cite{ther-1,ther-2,mag,sano2} by absorbing light selectively on the active surface. Taking advantage of these thermal or chemical gradients, self-phoretic processes become operational to produce mechanical force for propulsion.

Active particles exhibit fascinating non-equilibrium phenomena, like autonomous directed motion in the spacial periodic structures with broken inversion centre~\cite{JPCL,MSshort,ratchet,Olson}, giant drift velocity opposite to the applied force~\cite{GNM,BQ-1}, transient drifts toward fuel concentration gradients reminiscent of chemotactic motions~\cite{chemo-taxis,Lozano,Vuijk} etc. Further, active particles display an interesting collective phenomenon: motility-induced phase separation (MIPS)~\cite{Marchetti,Buttinoni,Redner,Cates,Bechinger,Golestanian,Dolai,DSR1,DSR2}. Through this mechanism, active particles coexist between two phases of different densities. Conspicuous transport features, unusual collective behaviors, and huge application potential motivate researchers to fully understand diffusion of active particles in crowded environments~\cite{Stewart,Gompper}, under various types of geometric constraints~\cite{entropic,Ao,fily,Caprini,Fily2,Jeyaram,JC-Wu,TDebnath,Ass-3b,Castro1,Ass-3a,Castro2,Ass-4,Ass-5} and fluid flow~\cite{Ramaswamy,Zöttl,Torney, Li1, Caprini2}.

In this paper, we restrict our attention to the driven transport of point-like self-propelled particles in a two dimensional array of symmetric obstacles [see Fig.~1 (a)].  In order to quantitatively interpret the simulation results, we use a simplified model of active Brownian particles in a highly viscous medium. We focus on situations where the impact of hydrodynamics, inertia inter-particle interactions, excluded volume, particle shape, spatial disorder, and nontrivial effects due to fluidics and chaos can safely be ignored. Further, we assume that the external drive is only coupled to the active colloidal particle without affecting the suspended fluid. 

 Earlier studies~\cite{obstacle-our,He1,Dagdug1} on transport of passive Brownian particles in the array of obstacles show that diffusion and mobility are considerably affected due to the trapping of driven particles against the obstacles. Pinning action and transverse diffusion-assisted depinning mechanisms make diffusivity and mobility a highly non-monotonic function of drive. Impacts of obstacles on diffusion of active particles have also been studied~\cite{Hamidreza, Pattanayak, Ao1}. However, these studies are restricted to unbiased situations.

Our study shows that due to the very nature of accumulation of the active particles on obstacles, a remarkable manifestation of geometric effects is observed in the driven transport. Emphasis has been given for 
 external bias's two directions --- parallel and diagonal to the principal array axis. For the both directions of drive, persistent length dependent depinning effects are signalled through excess diffusion peaks or sudden drops in diffusivity   and negative differential mobility. The persistent self-propulsion motion perpendicular to the drive helps particles in overcoming stagnation regions. However, trapping of active particles against obstacles has a considerably stronger impact on diffusivity and mobility in comparison to the passive Brownian particle. Further, the observed manifestations of trapping actions and the depinning phenomenon bear some similarities with transport traits of active matter on periodic substrates~\cite{Ass-1,Ass-2} and in the array of convection roll~\cite{Li-entropy,CPL}.

We organize  this paper as follows: Sec.~II presents
our model of active Brownian particles in the framework of the Langevin equation formalism, which is employed in our simulation code. We also discuss the significance of the relevant model parameters. Our main numerical results are
presented and interpreted with some analytic arguments in Sec.~III. In Sec.~III A, we analyse transport at the zero drive limit through  estimations of mean escape time from a channel compartment. Simulation results for parallel and diagonal directions of drive with respect to the principal array axis have been studied in Sec.~III B and III C, respectively.  Finally,  we summarize our results with some concluding remarks in Sec.~IV.

\section{Model} Let us consider an active particle diffusing in a highly viscous fluid contained in a 2D rectangular array of reflecting obstacles. The array of circular obstacles (disks) is characterized by the unit cell, $L_x \times L_y$ and obstacle radius $r_0$ [see Fig.1(a)]. Active particles have their own self-propulsion velocity $\vec{v}_0$ operated through some internal mechanism. For a self-propelled Janus particle, $\vec{v}_0$  is oriented parallel to a symmetry axis of the particle as illustrated in Fig.1(c). Fluctuations in fuel density or the intrinsic rotational diffusion result in random direction changes of the self-propulsion velocity. The particle is additionally driven by an homogeneous force $\vec{F}$ directed at an angle $\phi$ with respect to x-axis as depicted in Fig.1(a). From the consideration of immediate physical intuition, the overdamped dynamics of the active particle in the free space can be encoded by the following set of stochastic differential equations,
\begin{subequations}
\begin{eqnarray}
 \dot x &=& v_0\cos \theta +(F/\gamma) \cos \phi  +\xi_x(t),\label{Lan1}\\
 \dot y &=& v_0\sin\theta +(F/\gamma) \sin \phi  +\xi_y(t),\label{Lan2}\\
 \dot \theta &=& \xi_{\theta}(t),\label{Lan3}
\end{eqnarray}
\end{subequations}
The active particle's instantaneous position $(x,y)$ diffuses through the array of reflecting circular obstacles under the combined action of self-propulsion [$\vec{v_0}\equiv(v_0\cos\theta, \;v_0\sin\theta)$], external bias [$\vec{F}\;\equiv(F\cos\phi, \;F\sin\phi)$] and equilibrium thermal fluctuations [$\vec{\xi} \equiv (\xi_x, \; \xi_y)$]. The self-propulsion velocity with a constant modulus ($v_0$) is oriented at an angle $\theta$ with respect to the x-axis. Time evolution of $\theta$ due to rotational diffusion is modeled by a Wiener process Eq.(\ref{Lan3}) with $\langle \xi_{\theta}(t)\rangle=0$ and $\langle \xi_{\theta}(t)\xi_{\theta}(0)\rangle=2D_\theta\delta (t)$.  Where, amplitude of the rotational diffusion constant ($D_{\theta}$) depends on the viscosity ($\eta_v$) of the medium, temperature ($T$) and particle size. For a spherical particle of radius $a$, the rotational diffusion can be represented by Stokes-Einstein-Debye law, $D_{\theta} = k_BT/8\pi\eta_v a^3$. It should be noted that $D_{\theta}$ may contain contributions due to fuel density fluctuations that largely depend on the  internal mechanism of  self-propulsion.

The components of the self-propulsion force, ($v_0\cos\theta, v_0\sin\theta$), are alike to  the components of a 2D non-Gaussian
noise  with zero mean, $\langle
v_0\cos \theta\rangle=\langle
v_0\sin \theta\rangle=0$, and finite-time correlation functions,
$v_0^2 \langle \cos
\theta (t) \cos \theta (0) \rangle = v_0^2 \langle \sin \theta (t)\sin
\theta (0)\rangle =(v_0^2/2) e^{-|t|D_\theta}$. The correlation time, $\tau_\theta=1/D_\theta$ represents rotational relaxation time or persistence time for self-propulsion motion. The associated persistence length is defined as, $l_\theta = \tau_\theta v_0$. Self-propulsion of an active particle is fully characterized by two independent parameters, $\{v_0, \; D_\theta\}$  or $\{l_\theta, \; \tau_\theta\}$.

The last terms, $\xi_x(t)$ and $\xi_y(t)$ in the Eq.~(1a) and Eq.~(1b), respectively, represent independent thermal  fluctuations. They are modelled  by a Gaussian white noises with $\langle \xi_{q}(t)\rangle=0 $ and $\langle \xi_{q}(t)\xi_{q'}(0)\rangle=2D_0\delta (t)\delta_{qq'}$, where,  $q, q'=x,  y$.  The thermal noise strength $D_0$ is the measure of the translational diffusion of a free active particle in the bulk with $v_0 = 0$. Further, for passive colloidal particles both the rotational and translational diffusion are of only thermal origin. Thus, $D_0$ and $D_\theta$ are related as, $D_{\theta} = 3D_0/4 a^2$.  However, for an active particle, the mechanisms and origins of the translational and rotational diffusion may
not be the same. Thus, we consider $D_0$, $v_0$, and $\tau_\theta$ are independent model parameters. As noted earlier, for the sake
of simplicity, our study ignores particle-particle
collisions \cite{Buttinoni} and hydrodynamic impacts \cite{Ripoll}. This situation corresponds to a very dilute solution of active particle at low Reynold number. 

To explore transport properties of active particles we numerically  integrate coupled stochastic differential equations~(1a-1c) by means of a standard Milstein scheme \cite{Kloeden}. We assume that the obstacles are perfectly reflecting and the particle-obstacles collisions
are perfectly elastic. On collision with an obstacle, the active particle's instantaneous velocity ($\vec{\dot{r}}$) direction gets  reverted. To ensure numerical stability, we use a very short integration time step, $10^{-4}-10^{-5}$.  The mobility ($\mu$) and diffusivity ($D$) are defined as,
\begin{eqnarray}
 \mu(F) &=& \frac{1}{F}\lim_{t \rightarrow \infty} \frac{\langle[x(t)-x(0)]\rangle}{t} ,\label{Lan4}\\
 D(F) &=& \lim_{t \rightarrow \infty} \frac{[\langle x^2(t)\rangle-\langle x(t)\rangle^2]}{2t}. \label{Lan5}
\end{eqnarray}
Inspecting these quantifiers, we characterize transport features as well as the non-linear dynamical response of the active particles to the external force $\vec{F}$.

The simulation results shown in the Fig.~(2-6) are produced by ensemble averaging over $10^4 - 10^6$ trajectories depending upon the values of self-propulsion parameters and external bias. Further, transport quantifiers ($\mu$ and $D$) have been numerically estimated in the long time limit when effects due to transient processes and initial conditions die out. In our simulations, times and lengths are in seconds and microns, respectively.

\begin{figure}
\centering
\includegraphics[width=0.4\textwidth,height=0.45\textwidth]{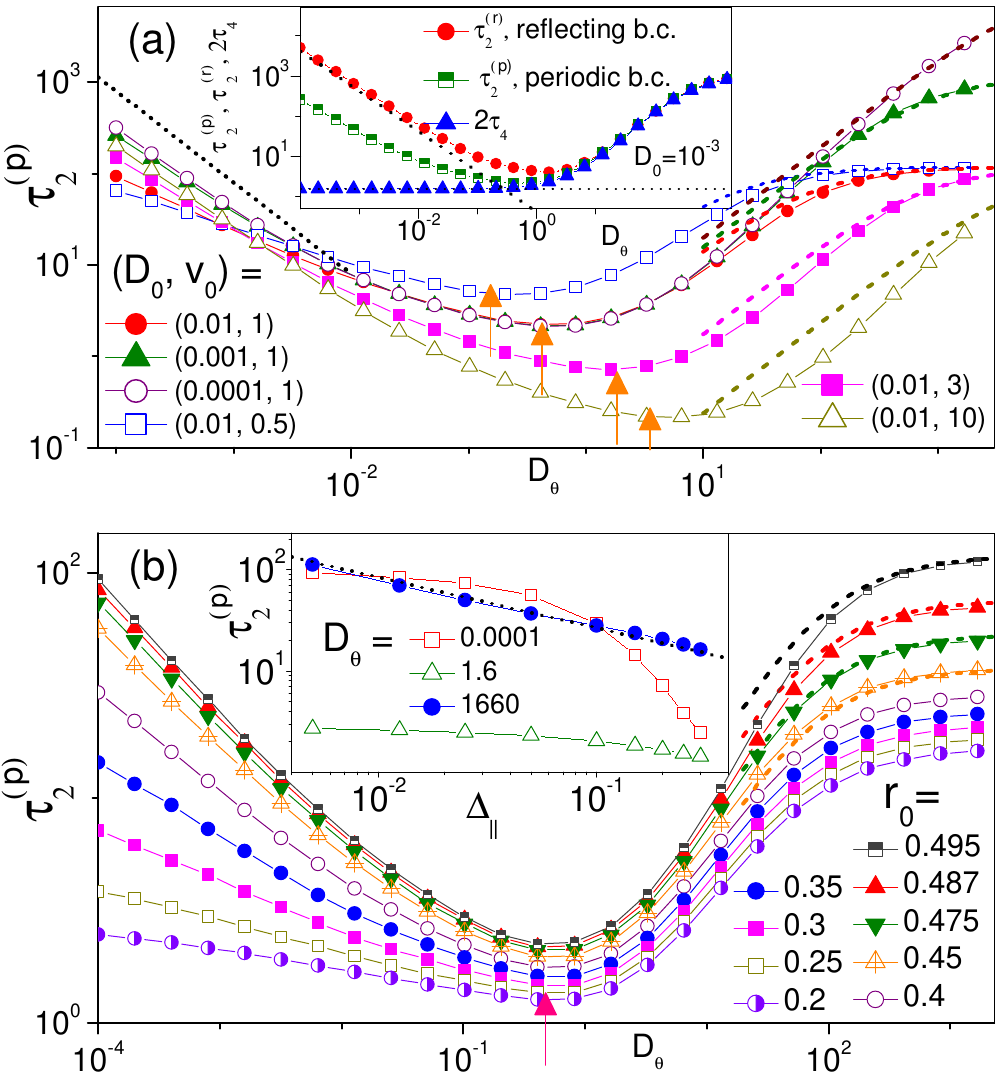}
\includegraphics[width=0.4\textwidth,height=0.225\textwidth]{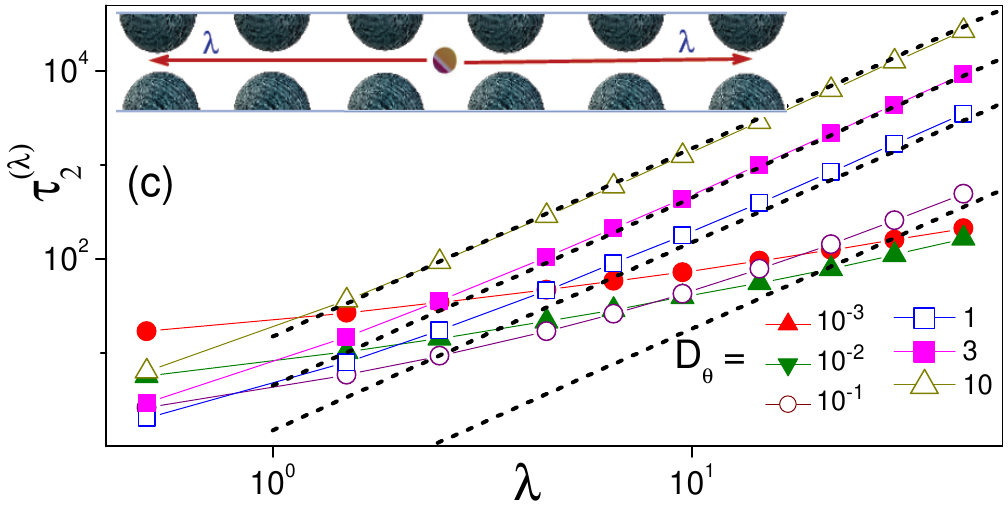}
\caption {(Color online) (a)  $\tau_2^{(p)} \; vs. \; D_\theta$ for different $v_0$ and $D_0$ (see legends). Dashed lines are analytic predictions [based on Eq.~(\ref{MFPT-1})] in the limit of very fast rotational dynamics.  The inset compares the variation of $\tau_2^{(p)},\; \tau_2^{(r)}$ and $2\tau_4$ with $D_\theta$.  The dotted line here indicates a lower bound of  $2\tau_4$ based on the Eq.~(\ref{mfpt-tau4}). The dashed line represents Eq.~(\ref{mfpt-2}).  
(b) $\tau_2^{(p)} \; vs. \; D_\theta $ for different obstacle radius $r_0$ (see legends). The inset depicts $\tau_2^{(p)} \; vs. \; \Delta_{\parallel}$. The dashed lines (both in the inset and the main panel) displays Eq.~(\ref{MFPT-1}). 
 (c)  $\tau_2^{(\lambda)} \; vs. \; \lambda $ with varying $D_\theta$ (see legends). Dashed lines represent Eq.~(\ref{mfpt-lambda}) with $\chi = 0.75$.  Simulation parameters (unless reported otherwise in the legends): $D_0 = 0.01, \; r_0 = 0.495,\; x_L=1, \; y_L=1, \; v_0 = 1$. Vertical arrows locate the minima using Eq.~(\ref{mfpt-min}).} 
\end{figure}  

\section{Results and Discussions}
To capture key transport features of driven active particles in the array of symmetric obstacles [as shown in Fig.1(a)], we produce simulation results with varying directions of applied bias $\vec{F}$ over the range $0 \to \pi/2$ (not shown). However, emphasis has been given on two limiting cases, namely, when the drive is parallel ($\phi = 0$) and diagonal [$\phi = \arctan(x_L/y_L)$] to the array axes. When force is directed along a lattice axis, spatial symmetry allows us to consider the array of obstacles as an array of symmetric corrugated channels parallel to $\vec {F}$ as depicted in the Fig.~3(a). We refer this type of periodic spatial structure as ``channel I". Another type of channel (referred as ``channel II" ) with a different topology emerges when $\vec{F}$ is directed at an angle $\phi = \arctan(x_L/y_L)$. In this situation, driven transport through the array can be regarded as transport through channel II as shown in the Fig.~1(a) and Fig.~5(a). For a rectangular lattice ($x_L \neq y_L$) with diagonal drive, the cross-section and period of a channel compartment are $2x_Ly_L/\sqrt{x_L^2+y_L^2}$ and $\sqrt{x_L^2+y_L^2}$, respectively. The transport features of active particles for parallel and diagonal biases have been analyzed separately. 

\subsection{Escape kinetics from channel compartments and transport in the zero drive limit} 
To better understand the transport mechanism in the zero forcing limit, we first explore escape kinetics of an active particle from a channel compartment. The transport quantifiers, diffusivity and mobility can directly be correlated  with the mean exit time.  Mean exit times through fours pores, $\tau_4$ , as well as, two opposite pores, $\tau_2$ of a channel compartment [see Fig.1(b)] have been numerically simulated as function of rotational diffusion constant with varying compartment geometry and self-propulsion velocity.  The active particle is injected in the middle of a channel compartment [see Fig.~1(b)] with a fully random orientation of $\vec{v_0}$ over the range [$0 \to 2\pi$].  To simulate $\tau_4$, we consider the particle is absorbed as  soon as it crosses the center of any one of the bottlenecks.  While estimating $\tau_2$, particles are absorbed at any two opposite pores and a periodic/reflecting boundary condition is assumed for exit through other two openings.   The mean exit time  with periodic and reflecting boundary are denoted as $\tau_2^{(p)}$ and $\tau_2^{(r)}$, respectively.  In Fig.~2(a,b), we depict variation of $\tau_2$ with $D_\theta$ over the range starting from  $l_\theta \gg  \{x_L, y_L\}$ to $l_\theta \ll  \{\Delta_{\perp},\Delta_{\parallel}\}$. All $\tau_2^{(p)} \; vs.\;D_\theta$ plots pass through a minimum located at,
\begin{eqnarray} 
 D_{\theta}^{min}=  {v_0}/{2x_L} \label{mfpt-min}
 \end{eqnarray}
This condition corresponds to the matching of compartment length with half of the persistence length. 

In the inset of Fig.~2(a), we compare the variation of  $\tau_2^{(p)}$, $\tau_2^{(r)}$ and  $2\tau_4$ as a function of $D_\theta$. 
Similar to  $\tau_2^{(p)}$,  $\tau_2^{(r)}$ exhibits a minimum as a function of $D_\theta$. However, position of minimum shifts to higher $D_\theta$ and $(\tau_2^{(r)})_{min}>(\tau_2^{(p)})_{min}$,  as particles require more rotation to exit under reflecting boundary condition. Here, the minimum  is located at, $D_{\theta}^{min} \sim  {v_0}/{x_L}$. This relation amounts to matching of self-propulsion length with the compartment size.    It appears from simulation results [in Fig.2(a)] that the relation, $\tau_2^{(p)}=\tau_2^{(r)}=2\tau_4$, is restricted for $l_\theta \ll  \{x_L, y_L\}$.   This observations, along with other simulation results in Fig.~2 show that the escape mechanism for fast rotational diffusion is apparently different from its opposite limit of slow rotational relaxation. We discuss these two regimes separately. Further, our analysis mainly focuses on $\tau_2^{(p)}$ as it is directly connected with zero-forcing limit transport through the array of obstacles.    
\begin{figure}
\centering
\includegraphics[width=0.4\textwidth,height=0.7\textwidth]{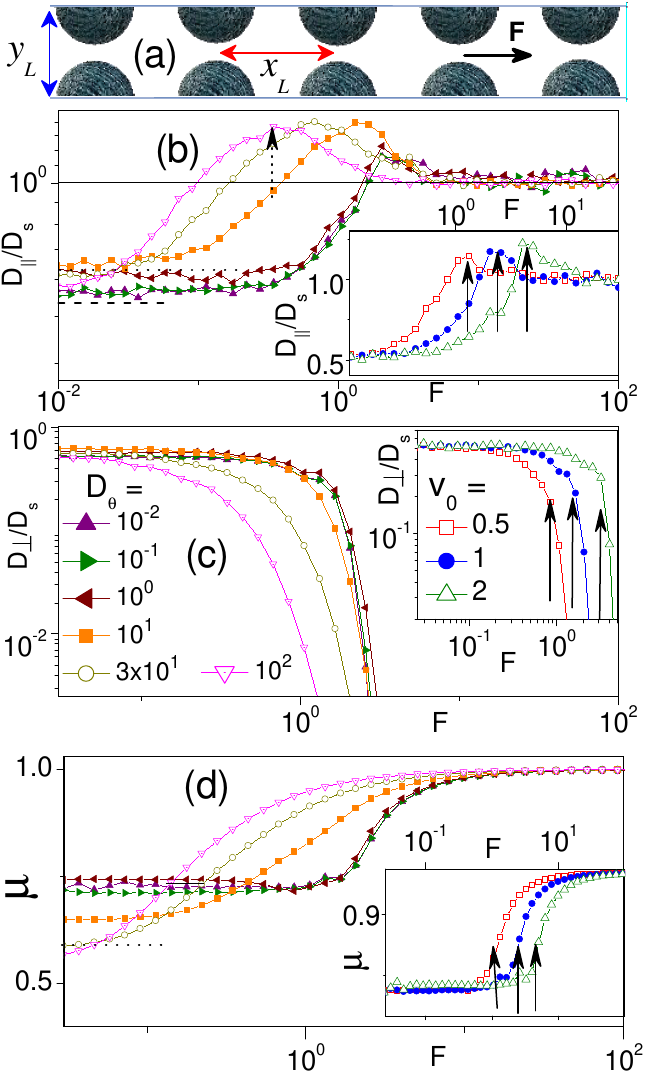}
\caption {(Color online) (a) Schematic of the corrugated channel (channel I) for $\vec{F}$ parallel to the horizontal axis.  $x_L$ and $y_L$ represent channel periodicity and width, respectively. Panel (b) and (c) depict $D_\parallel \; vs. \; F$  and $D_\perp \; vs. \; F$ for different  $D_\theta$ as shown in the legends. Panel (d) shows mobility ($\mu$) versus $F$. Simulation parameters (unless reported otherwise in the legends): $v_0 = 1.0, \; D_0 = 0.01, \; r_0 = 0.45,\; x_L=y_L=1$. Insets of the panels (b), (c) and (d), represent $D_\parallel$ versus $F$,  $D_\perp$ versus $F$   and $\mu$ versus $F$, respectively, for different $v_0$ (see legends). Other parameters are the same as the  main panels but $D_\theta = 0.1$ .  Dotted lines depict limiting values of $D(F)$ and $\mu(F)$ for $F\rightarrow 0$ and $l_\theta \ll x_L$, estimated using Eq.~(5--8). The solid line indicates the asymptote, $D_\parallel (\infty)/D_s =\mu (\infty) = 1$. For $l_\theta \gg x_L$, estimation of $D(0)/D_s$ based on Eq.~(\ref{diff-chi}) is displayed by the dashed line.   Vertical arrows indicate $D_\parallel$ peaks and corresponding transition points in $D_\perp$ and $\mu$. Note that dotted and solid arrows correspond to the depinning threshold estimated based on the Eq.(\ref{dp-2}) and Eq.(\ref{dp-3}), respectively.  }
\end{figure}

\begin{figure}
\centering
\includegraphics[width=0.4\textwidth,height=0.65\textwidth]{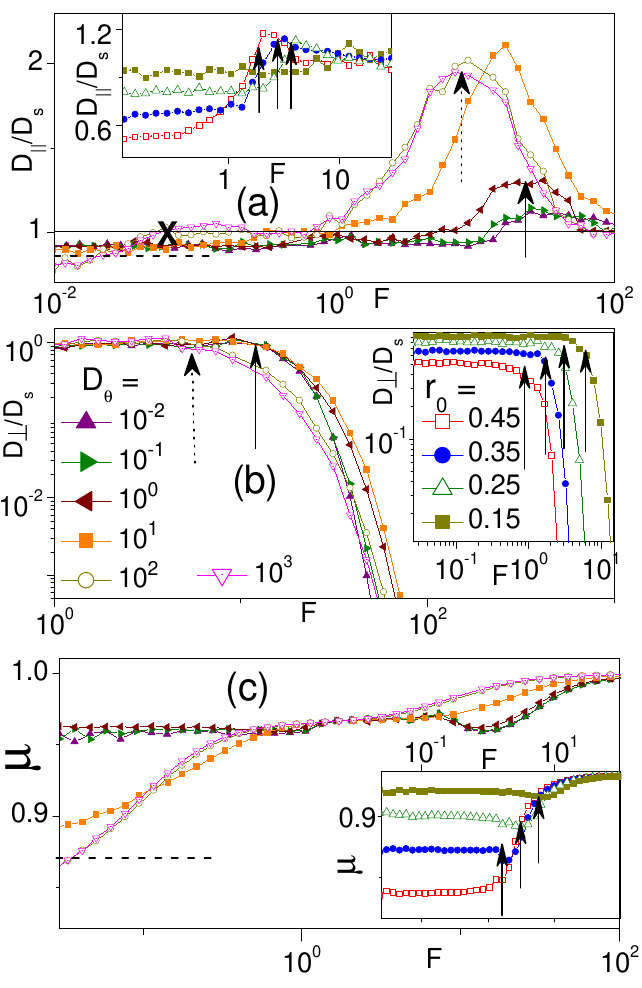}
\caption {(Color online) Transport quantifiers for driven transport in channel I.   Panels (a), (b) and (c) depict $D_\parallel \; vs. \; F$, $D_\perp \; vs. \; F$, and $\mu \; vs. \; F$, respectively, for different $D_\theta$ (shown in the legends) and $x_L \neq y_L$.  Simulation parameters (unless reported otherwise in the legends): $v_0 =1.0, \; D_0 = 0.03, \; r_0 = 0.4,\; x_L=8, \; y_L=1$. Insets of the all panels  represent similar plots as the corresponding main panels but for different obstacle radius. Parameters for the inset:  $D_0 = 0.01, \; D_\theta = 0.1,\; x_L= y_L=1$. Dashed lines depict limiting values of $D(F)$ and $\mu$ for $F\rightarrow 0$ and $l_\theta \ll x_L$, estimated using Eq.~(5, 8). The solid line indicates $D_\parallel (\infty)/D_s =\mu (\infty) = 1$. Analytic estimates of depinning thresholds $F^{(I)}_{D_1}$ , $F^{(I)}_{D_2}$ and $F^{(I)}_{D_3}$ are indicated  by  cross, dotted arrows and solid arrows, respectively.}
\end{figure}

\subsubsection{Fast rotational relaxation}  When the rotational relaxation is so fast  that the persistent length of self-propulsion is even smaller than the channel bottlenecks [$\Delta_{\parallel}=y_L/2 - r_0,\; \Delta_{\perp}=x_L/2 - r_0$], the motion of the active particles in a channel compartment can be considered as uncorrelated. This limiting situation allows us to assume that  the system is in equilibrium condition at an effective temperature, $T_{eff} = \gamma (D_\theta D_0+v_0^2)/D_\theta k_B$ . In this region,  $\tau_2^{(p)}$ is inversely related to $\sqrt{\Delta_\parallel}$ and $T_{eff}$. Following ref.~\cite{Holcman,obstacle-our}, we obtain an expression of $\tau_2^{(p)}$ as,
\begin{eqnarray} 
 \tau_2^{(p)}= \left(1-\frac{\pi}{4}\right) \frac{\pi x_L^2 D_\theta}{4\left(2D_0 D_\theta+v_0^2\right)}\sqrt{\frac{x_L}{\Delta_\parallel}} \label{MFPT-1}
 \end{eqnarray}
Estimations of $\tau_2^{(p)}$ based on the above equation are well accord with the simulation data presented in the Fig.~2 for $v_0/D_\theta > \Delta_{\parallel}$. It should be noted that both  $\tau_2^{(r)}$ and  $2\tau_4$ also follow the Eq.~(\ref{MFPT-1}). 

For this regime of fast rotational dynamics, diffusivity and mobility in the zero forcing limit are connected through Einstein relation,     
\begin{eqnarray} 
 D(0)/D_s =\mu(0). \label{Dif-2}
 \end{eqnarray} 
Where, the free space diffusivity of an active particle, 
\begin{eqnarray} 
 D_s=D_0+v_0^2/2D_\theta \label{Dif-1}.
 \end{eqnarray}
Simulation results in Fig.~3 satisfy the relation (\ref{Dif-2}) for $D_\theta = 100$. 

Moreover, $D(0)$  and $\mu(0)$ can be estimated from the knowledge of mean exit time from a channel compartment $\tau_2^{(p)}$.  The mean waiting time in a compartment of the channel I is $2\tau_2^{(p)}$. Assuming particle's transport through the channel as random work in a 1D lattice, diffusivity can be estimated as,   
\begin{eqnarray} 
 D(0)={x_L^2}/{4\tau_2^{(p)}} \label{Dif-3}
 \end{eqnarray}
Estimates of $D(0)$ and $\mu(0)$ based on the Eqs.~(\ref{Dif-2}-\ref{Dif-3}) are indicated by dotted horizontal lines in the Fig.~3. Simulation data are very close to these predictions.

\subsubsection{Slow Rotational Relaxation} 
For the very slow relaxation time and $v_0$ is much larger than the strength of the thermal fluctuations, both $\tau_2^{(p)}$ and  $\tau_2^{(r)}$ are inversely related to $D_\theta$. This attribute to the fact that self-propulsion pushes the particle against the obstacle and its movement leading to exit requires rotational diffusion of the particle over an appropriate angle.  We first get an approximate estimation of $\tau_2^{(r)}$ based on the following consideration.  
Recall that self-propulsion velocity is uniformly distributed over the angle 0 to $2\pi$. Escape of a trajectory requires rotation of particles through an angle in between 0 to $\pi$ so that self-propulsion velocity becomes directed to channel bottlenecks. Thus,  the exit problem is basically reduced to 1D diffusion where particles are uniformly distributed between two absorbing points at 0 and $\pi$. Based on this reasoning the mean exit time is given by, 
\begin{eqnarray} 
 \tau_{2}^{(r)}=  {\pi^2}/{24 D_\theta} \label{mfpt-2}
 \end{eqnarray} 
Estimations based on this equation [shown in the inset of Fig.~2(a)] are well accord with simulation results. However, this equation is not valid for exit through two opposite pore with periodic boundary along the transverse direction. Nevertheless, an inverse relation, $\tau_2^{(p)} = A_p /D_\theta$, holds with proportionality constant $A_p \ll \pi^2/24 $.  This implies that the particles manage to exit even with substantially less rotation of self-propulsion velocity direction. Due to the combined effects of periodic boundary and the focusing action of $\vec{v_0}$, particles smoothly glide to the bottlenecks as soon as self-propulsion pushes them against the round shaped obstacles. Thus, the exit process gets much faster than estimated in Eq.~(\ref{mfpt-2}).  

On the other hand, the mean escape time through four openings is insensitive to $D_\theta$ in the region of slow rotational dynamics. 
 The lower bound of $\tau_{4}$ is the time required to drift over the length $x_L/2$ with an average velocity $2v_0 /\pi$. This produces,
\begin{eqnarray}
 \tau_{4}=  {x_L\pi}/{4v_0} \label{mfpt-tau4}
 \end{eqnarray}
This estimation is marked by a dotted line in the inset of Fig.~2(a). Simulation data exactly fall on the dotted lines. This results corroborate our assertion  that exit from a compartment with four openings does not require rotational diffusion. Only focusing action of self-propulsion forces is enough to {\it escort} the particles to the exit windows.

Figures~(3-6) show that transport features start remarkably changing with growing self-propulsion length beyond $\Delta_{\parallel}$. In the relatively low drive regions ($F < v_0$), the diffusivity and mobility are insensitive to the self-propulsion parameters as well as $F$ when $l_\theta > \{ x_L,y_L\}$.  
Contrary to the opposite limit, $l_\theta \ll \{ x_L,y_L\}$, assumption of equilibrium-like state with an effective temperature is not valid here. Further, for $l_\theta \geq x_L$,  a number of successive compartment crossing events become correlated. Thus, both the  Eq.~(\ref{Dif-3}) and Eq.~(\ref{Dif-2}) lose their validity.

 To connect the exit events with diffusivity for the slow rotational  limit, we estimate mean first passage time $\tau_2^{(\lambda)}$  by setting absorbing points at a distance $\lambda$ [see inset of Fig.2(c)] which is much larger than $l_\theta$. Here, we assume periodic boundary conditions along the perpendicular direction of the channel axis.  Figure 2(c) depicts the variation of  $\tau_2^{(\lambda)}$  with $\lambda$.    When  $\lambda$ is large enough, consecutive events of crossing over the length $\lambda$ occurs totally   uncorrelatedly, as a result, $\tau_2^{(\lambda)}$ becomes directly proportional to $\lambda^2$. Simulation results fit well with following empirical relation,   
\begin{eqnarray}
 \tau_{2}^{(\lambda)}=  \frac{2 \chi \lambda^2 D_\theta}{v_0^2}, \label{mfpt-lambda}
 \end{eqnarray}      
where, $\chi$ is a dimensionless parameter which depends only on the channel geometry. Note that validity of this equation requires $v_0^2/2D_\theta \gg D_0$. Now, suitably setting the absorbing points one can estimate the diffusivity in the zero forcing limit by substituting $x_L$ by $\lambda$ and $\tau_2^{(p)}$ by $\tau_2^{(\lambda)}$ in Eq.~(\ref{Dif-3}). This leads to, 
\begin{eqnarray}
D(0)/D_s \sim 1/8\chi. \label{diff-chi}
\end{eqnarray} 
 Estimation  based on this equation indicated in Fig.~3(b) well collaborates simulation results. Further, Eq.~(\ref{diff-chi}) is accord with the fact that $D(0)/D_s$ is insensitive to self-propulsion velocity and rotational relaxation as long as $l_\theta \geq x_L$ (see Fig.~3-4). It should be noted that in the zero drive limit, $D_{\parallel}(0) = D_{\perp}(0)$, irrespective of the direction of  the force ${\vec F}$. 

\subsection{Transport with parallel drive } 
Recall that transport with a constant drive  parallel to the array axis amounts to driven transport through an array of connecting channels of type I [see Fig.~1(a) and Fig.3(a)]. Where, the applied force $\vec {F}$ is directed along the channel axis [Fig.3(a)]. We calculate diffusivity parallel to the drive, $D_{\parallel}(F)$, as well as,  its perpendicular direction $D_{\perp}(F)$.  In Fig.~3-4, we represent some significant simulation results which capture key features of driven transport of active particles through channel I.  Depending upon the amplitude of $F$, $D_{\parallel}(F)$ can be enhanced or suppressed.  In the very low forcing region diffusion and mobility of active particles get noticeably suppressed. On the other hand, for very large drives $D_\parallel (F)$ and $\mu (F)$ reach their free space values. In the intermediate region of forcing, transport occurs with excess diffusion.

 Here, $D_\parallel(F) \; vs.\; F$ plots exhibit single or double  peaks depending upon the amplitude of $\l_\theta$ and channel compartment geometry.  The peaks positions, as well as, their heights depend on the self-propulsion parameters, thermal diffusion and the channel shape. Transverse diffusivity, $D_\perp(F)$, remains insensitive to $F$ up to some threshold value. Beyond that threshold, $D_\perp (F)$  exponentially decays very fast.
Followed by a subtle minimum a sharp increase in $\mu (F) \;  vs.  \; F$ indicates the presence of this threshold driving strength.  These interesting observations are associated with some pining actions and subsequent depinning processes.
Note that here trapping or pinning  of an active particle refers to a state of stagnation with limited movement due to geometric constraints and the depinning process is the transition between  pinned and moving phases~\cite{Marchesoni-RMP}.

 It appears from simulation results in Fig.~(3-4) that intriguing transport features  for very slow rotational relaxation  are noticeably different from the opposite limit of fast rotational dynamics ($l_\theta \ll \{x_L,\; y_L\}$). It should be noted that the self-propulsion effect dominates in the transport of the particles as long as $\l_\theta \geq  \{x_L,\; y_L\}$ and $v_0^2/2D_\theta > D_0$.  In the following we analyse  depinning mechanism in the two distinct region of rotational dynamics. 
 
\subsubsection{Depinning mechanism for fast rotational relaxation}  
For $x_L = y_L$, with increasing $F$,  $D_\parallel(F)/D_s$ grows to a maximum then gradually decays to unity. An additional peak emerges for a rectangular channel compartment with $x_L \gg y_L$ in the low forcing region. Similar to passive colloidal particles~\cite{obstacle-our}, two excess diffusion peaks are associated with two different depinning mechanisms. For low drives, the possibility of particles getting struck against the obstacle is most when the transverse diffusion time over the  distance $r_0$ matches with the time to drift the same length along the channel axis.  Thus the corresponding depinning threshold is given by, 
\begin{eqnarray} 
 F_{D_1}^{(I)}=  \frac{ \left(2D_0 D_\theta+v_0^2\right)}{D_\theta r_0} \label{dp-1}
 \end{eqnarray} 
On the other hand, for the relatively high forcing region, the particle trapping in the array of the obstacle becomes most effective when transverse diffusion time over the half-width of the reduced channel width ($y_L/4$) becomes comparable to the drift time to travel  the separating distance ($x_L-r_0$) between a center of bottleneck and the nearest pair of obstacles. Comparing these two time scales we obtain the second depinning threshold, 
 \begin{eqnarray} 
 F_{D_2}^{(I)}=  \frac{ 64D_\theta \left(x_L-r_0\right)}{y_L^2\left(2D_0 D_\theta+v_0^2\right)} \label{dp-2}
 \end{eqnarray} 
The expressions (\ref{dp-1}-\ref{dp-2}) predicting depinning thresholds are generalization of Eqs.(5-6) in Ref.~\cite{obstacle-our} for active Brownian motion in the fast rotational diffusion limit. Estimation of these  threshold forces, $F_{D_1}^{(I)}$ and  $F_{D_2}^{(I)}$   using Eqs.~(\ref{dp-1}-\ref{dp-2}) for $D_\theta = 100$ are indicated by cross and dotted vertical arrows, respectively [see  the Fig.~3(b) and Fig.~4(a)]. The predicted threshold values clearly locate the positions of excess diffusion peaks.

The notion of excess diffusion due to depinning mechanism was introduced by Costantini {\it et.~al}~\cite{Costantini-EPL}. They studied passive Brownian diffusion in a tilted washboard potential both in the overdamped and underdamped limits. In the present context, particles can propel themselves and  periodic structure is due to an array of obstacles. Thus, the depinning mechanism here becomes considerably different. 
 
\subsubsection{Depinning mechanism for slow rotational relaxation}   
 
With decreasing $D_\theta$, the peak position in a $D_{\parallel} \; vs.\; F$ gradually moves to the larger driving strength. For $l_\theta \geq x_L$ and a given $v_0$, both the peak position and its height are insensitive to $D_\theta$. In this self-propulsion dominated region, the excess diffusion peak  is much weaker. 
 The depinning process here is a result of interplay between self-propulsion and drive $F$. Depinning threshold here can be determined by comparing transverse drift time to cross obstacle radius, $\tau_{\perp}=r_0/v_0$ with the longitudinal drift time for compartment crossing $\tau_{\parallel}=x_L/F$. Associated depinning threshold is given by,
\begin{eqnarray} 
F_{D_3}^{(I)}=  \frac{x_L v_0}{r_0}, \label{dp-3}
\end{eqnarray} 
The positions of diffusion peaks, sudden jumps of non-linear mobility $\mu(F)$, and abrupt drop in transverse diffusion correspond to this depinning threshold. The positions of $F_{D_3}^{(I)}$ indicated by  solid vertical arrows [see Fig.~3 and Fig.~4] fairly agree with simulation results.

Unlike the thermal fluctuations dominated regions, here, the detention time against the obstacle is much lower. As the strong self-propulsion removes the particle from the stagnation area through its focusing action. This leads to suppression of the excess diffusion peaks. Further, only one diffusion peak is noticeable even for $ x_L > y_L$.

\begin{figure}
\centering
\includegraphics[width=0.4\textwidth,height=0.7\textwidth]{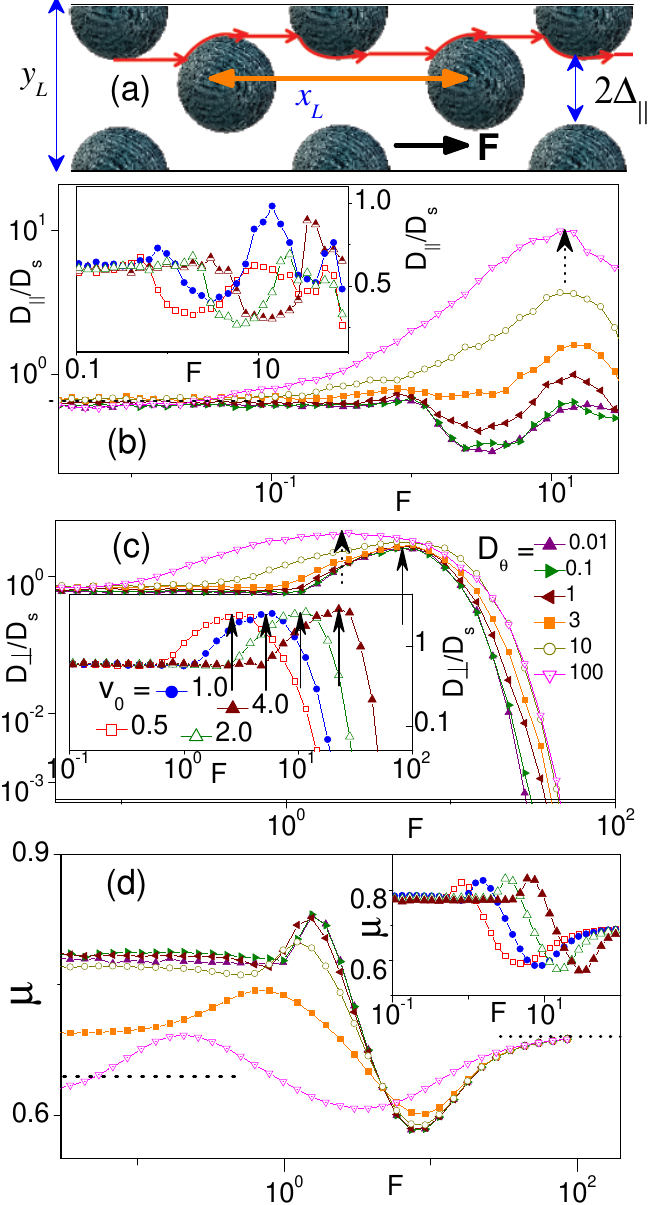}
\caption {(Color online) (a) Schematic of channel II. Here, $x_L$, $y_L$, and  $2\Delta_\parallel$ represent channel periodicity, width, and bottleneck, respectively. The red connecting arrows depict a particle trajectory guided by focusing/funnelling action of the drive in the limit $F \rightarrow \infty $.    Panel (b) and (c) depict $D_\parallel$ and $D_\perp$ versus $F$  for different  $D_\theta$ as shown in the legends. Panel (d) shows $\mu \; vs. \; F$. Simulation parameters (unless reported otherwise in the legends): $v_0 = 1, \; D_0 = 0.01, \; r_0 = 0.4,\; x_L=2,\; y_L=1$. Insets of the panels (b), (c) and (d), represent $D_\parallel \; vs. \; F$,  $D_\perp \; vs. \; F$   and $\mu \; vs. \; F$, respectively, for different $v_0$ (see legends). Other parameters are the same as the  main panels but $D_\theta = 0.1$. Dotted lines depict limiting values of $D(F)$ and $\mu(F)$ for $F\rightarrow 0$ and $l_\theta \ll x_L$, estimated using Eq.~(5 -- 8). The dashed line indicates the asymptote, $\mu (\infty)$ based on the Eq.~(\ref{mu-infinity} --\ref{t-infinity}).   Note that dotted and solid arrows correspond to the depinning threshold estimated based on the Eq.(\ref{dp4}) and Eq.(\ref{dp5}), respectively. }
\end{figure}

\begin{figure}
\centering
\includegraphics[width=0.4\textwidth,height=0.7\textwidth]{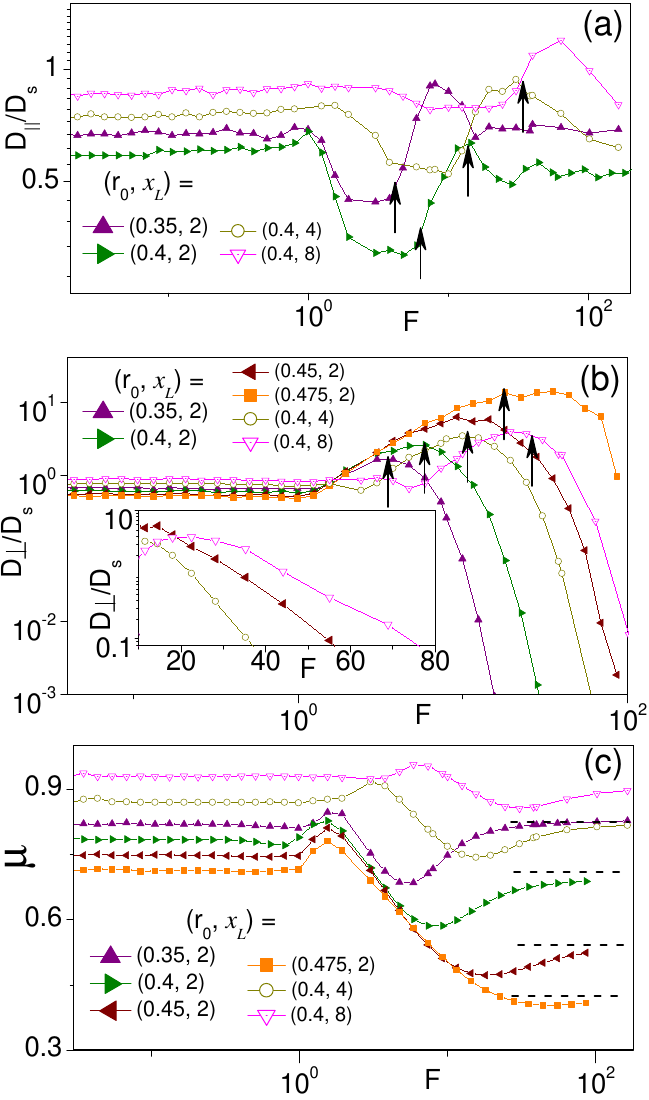}
\caption {(Color online) Simulation results for driven transport in Channel II.  Panel: (a), (b) and (b), depict $D_\parallel \; vs. \; F$, $D_\perp \; vs. \; F$, and $\mu \; vs. \; F$,  respectively, for different channel geometry (see legends).   Simulation parameters (unless reported otherwise in the legends): $v_0=1.0, \; D_\theta =0, \; D_0 = 0.01, \; r_0 = 0.4,\; x_L=2,\; y_L=1$. Dotted lines depict limiting values  $\mu$ for $F \rightarrow \infty$ according to Eqs.~(\ref{mu-infinity}-\ref{t-infinity}). Vertical arrows indicate depinning threshold estimated based on  Eq.~(\ref{dp5}). Inset of the panel (b) represents  $D_\perp \; vs. \; F$ (in semi-log scale) showing exponential decaying tail.
 }
\end{figure}

\subsection{Transport with diagonal drive}
Figures 5 and 6 show mobility and diffusivity [both, $D_\parallel(F)$ and $D_\perp(F)$] of a driven active particle in channel II as a function of the driving strength with varying self-propulsion parameters and channel geometries. Reduced channel II has off axial bottlenecks located alternatively against top and bottom. As a result, active particles moving through this channel are  more sensible to trapping action. Based on careful observation of the simulation results, notable transport features are listed below:

 (i) Similar to channel I, for fast rotational dynamics,    $D_\parallel (F) \; vs. \;  F$ plots  exhibit one or more excess diffusion peaks irrespective of channel geometry. However, obstacles here significantly increase the dispersion  in comparison to the parallelly driven particles.  Diffusion peaks position and their heights are very sensitive to the channel parameters  $\Delta_{\parallel}$ and $r_0$.  On the other hand, for slow rotational dynamics, over a very long drive range the diffusion along the channel axis is notably inhibited. Most remarkably,  diffusion gets minimized at some threshold driving amplitude.

 (ii) As soon as the drive grows larger than the self-propulsion force, transverse diffusion first significantly increases with increasing  $F$, however, followed by a maximum  $D_\perp(F)$ decays exponentially [see inset of Fig.~6(b)]. The self-propulsion parameters, $D_0$, and the channel geometry affect the associated decay constant.  Excess  transverse   diffusion peak is not observed for transport with parallel drive with respective array axes.

 (iii) For $l_\theta \geq x_L$ and $v_0^2/2D_\theta \gg D_0$,  the transport quantifiers, $D_\parallel(F)$,  $D_\perp(F)$ and $\mu(F)$, are insensitive to the drive as long as $F < v_0$. Sudden changes are observed when the driving strength wins over the self-propulsion.

  (iv) Here, $\mu(F)$ is a highly non-monotonic function and features a negative amplitude of its differential with respect to $F$. When $F$ grows larger than $v_0$, mobility first increases to a maximum then decreases with a power law $\mu(F) \sim 1/F^{\alpha} $ up to a certain driving strength. Finally, $\mu(F)$ grows again to an asymptotic value.  The amplitude of negative differential mobility (proportional to $\alpha$) of an active particle is much larger than the passive ones. For the simulation parameters corresponding to Fig.5(d), values of $\alpha$ for active and passive particles are about 0.25 and 0.06, respectively.

    (v) In the asymptotic limit, $F \rightarrow \infty$, the mobility $\mu(F)$ depends only on the channel geometry and is insensitive to the self-propulsion parameters and thermal fluctuations. However, the asymptote for diffusion along the channel axis behaves differently for fast and slow rotational dynamics. Nevertheless, for the both limits: $l_\theta \gg \{x_L,\; y_L\}$, and $l_\theta \ll \{x_L,\; y_L\}$,   $D_\parallel(\infty)/D_s$ becomes insensitive to the self-propulsion and thermal diffusion.
    
\subsubsection{Depinning mechanism for diagonal drive}     
All these features are restricted for $\Delta_{\parallel} < y_L/4$. As noted in the context of transport in channel I, impact of pinning action and depinning mechanism depends largely on the self-propulsion length. For $l_\theta \ll \Delta_{\parallel}$, active particle's motion can be considered as uncorrected Brownian motion with diffusivity $D_s$. Here, along with the transverse diffusion, depinning action is assisted by focusing action of ${\vec F}$. Due to funneling action particles move through a meandering path as shown in the Fig.~5(a). This action of drive becomes significant as long as the drift time ($\tau_\parallel$) to cover a distance from the centre of the bottleneck to the opposing obstacle matches with transverse diffusion time ($\tau_\perp$) for half of the bottleneck. These associated drift and diffusion times are estimated as,
\begin{eqnarray} 
\tau_\perp = \frac{8D_\theta \Delta_\parallel^2}{ 2D_\theta D_0+v_0^2 }, \;\;\;\tau_\parallel = \frac{x_L-2r_0 }{2F} \label{drift-diff}
\end{eqnarray}
 The corresponding depinning threshold is given by,
 \begin{eqnarray} 
F_{D_1}^{(II)} = \frac{4D_s(x_L-2r_0)}{ \Delta_\parallel^2 } \label{dp4}
\end{eqnarray}
This depinning threshold estimates the position of the excess diffusion peak in $D_{\parallel}(F) \; vs. \; F$ in the fast rotational limit. In Fig.~5(b), the peak position for $D_\theta = 100$ is indicated by a vertical arrow well corroborates simulation results.

However, in the self-propulsion dominated regions, $l_\theta > {x_L}$, transport quantifiers vary in a rather complicated way. As soon as $F$ surpasses $v_0$, diffusivity and mobility first begin to increase. Followed by a subtle maximum both $D_{\parallel}(F)$ and $\mu(F)$ start getting suppressed and eventually at some driving strength they get minimized. With further increasing $F$, mobility enhances to a saturation and an excess diffusion peak emerges. On the other hand, the variation of transverse diffusivity $D_{\perp}$ is much simpler. It exhibits only a very robust maximum. Interestingly, for a certain region of $F$, the interplay between self-propulsion and drive leads to suppression of both $D_{\parallel}(F)$ and $\mu(F)$, while it enhances transverse diffusivity. This implies that the obstacle restricts active particle's movement parallel to the drive, however, facilitates along the transverse direction.   These features in the self-propulsion dominated region can largely be understood based on the following considerations.

For $F>v_0$, driven active swimmers move through a meandering path [see Fig.~5(a)] with the assistance of focusing action of drive. However, this motion can be maladjusted due to self-propulsion motion along the transverse direction. The focusing action of the drive gets hampered considerably when the particle moves a distance $\Delta_\parallel$ along transverse direction faster than displacement of length $x_L/2-r_0$ along the channel axis. The  associated self-propulsion assisted transverse drift time is, $\tau_\perp \sim \Delta_\parallel/v_0$. This estimation of $\tau_\perp$ along with $\tau_\parallel$ in Eq.~(\ref{drift-diff}) yields following depinning threshold, 
\begin{eqnarray} 
F_{D_2}^{(II)} = \frac{v_0(x_L-2r_0)}{2 \Delta_\parallel } \label{dp5}
\end{eqnarray}
This threshold corresponds to the position of excess diffusion peak in $D_\perp(F)$, and inception of rising branches in $D_\parallel(F)$ and $\mu(F)$ followed by the flat minima.  Based on Eq.~(\ref{dp5}) the estimated peaks position in $D_\perp(F)\; vs. \; F$   [indicated by solid vertical arrows in Fig.~(5-6)] are fairly consistent with simulation data. Further, numerical results show that the minimum and the peaks are located at $F_{D_2}^{(II)}/2$ and $2F_{D_2}^{(II)}$, respectively.

To this end, the pattern of emerging dips, peaks and abrupt changes in the transport quantifiers attributed to the underlying pinning actions largely depends on the array geometry in addition to the direction of the drive. Irrespective of self-propulsion properties and the direction of the drive, we observe two separate depinning mechanisms which become apparent with increasing the ratio $x_L/y_L$.  In the fast rotational diffusion limit, the obstacles array geometry dependence  of  dips and peaks is expected to be similar as reported in the context of  kinetically locked-in colloidal transport in an array of optical tweezers~\cite{Ass-7}  and driven vortex lattices with periodic pinning~\cite{Ass-6}.  

\subsubsection{Transport for very large driving amplitude } 
It is apparent from Fig.~5 and Fig.~6, that in the large forcing limit both $D_{\parallel}(F)$ and $\mu(F)$ approach to some asymptotic value. Analytic estimate of asymptote in longitudinal diffusion $D_{\parallel}(\infty)$ is a formidable task. However, it appears from simulation data in the both fast and slow rotational diffusion limits,  $D_{\parallel}(\infty)$ is insensitive to the self-propulsion parameters as well as the thermal diffusion. Only the channel geometry determines the diffusivity. However, in the intermediate region self-propulsion length, $D_{\parallel}(\infty)$ depends on the self-propulsion parameters. The numerical data for very fast rotational dynamics can be reduced with the empirical relation \cite{obstacle-our,Borromeo},  $D_{\parallel}(\infty)/D_s=x_L r_0/2y_L \Delta_\parallel$.

The asymptote $\mu{(\infty)}$ can be estimated from the knowledge of mean exit time $\langle \tau_\infty \rangle$ from a channel compartment in the limit $F \rightarrow \infty$. In this limit,  mobility can be expressed as~\cite{Cox,Borromeo},    

\begin{eqnarray} 
\mu({\infty}) = \frac{x_L}{F \langle \tau_\infty \rangle}, 
 \label{mu-infinity}
\end{eqnarray}

We also calculate $\mu({\infty})$ through direct numerical simulation  of $\langle \tau_\infty \rangle$.  This estimation, as well as numerical data in Fig.~5(d) show that $\mu({\infty})$ is insensitive to $D_\theta$, $v_0$ and $D_0$. Only the channel parameters $x_L$, $y_L$ and $r_0$ determine the asymptotic value of the mobility. An exact analytic estimation of $\mu({\infty})$ can be obtained using expression of mean first passage time~\cite{obstacle-our},  
\begin{eqnarray} 
\langle \tau_\infty \rangle = \frac{x_L}{F} \left[1-\frac{\rho y_L}{x_L} +\frac{r_0}{x_L}\ln\left(\frac{2r_0+\rho y_L }{2r_0-\rho y_L }\right) \right], \label{t-infinity}
\end{eqnarray}
where, $\rho=\sqrt{2r_0/y_L-1}$. Estimations of $\mu(\infty)$ using Eq.~(\ref{mu-infinity}) along with Eq.~(\ref{t-infinity}) are indicated by dashed lines in Fig.5~(d) and Fig.~6(c).

We conclude this section with a comment on $D_{\perp} (\infty)$. When $F$ is much stronger than both the thermal fluctuations as well as self-propulsion, due to focusing action of the drive particle follows a meandering path  in Channel II [see Fig.~5(a)]. In this limit, the drive guides the active swimmer to move along the eccentric channel and compartment crossing along the transverse directions becomes a very rare event. Thus, $D_{\perp} (F)$ drops to zero as $F \rightarrow \infty$.
\newline
\newline


\section{Conclusions}     
We explore transport features of active particles in the 2D  array of circular obstacles. For parallel and diagonal direction of the external bias (with respect to the array axis) detailed analysis has been performed. Our simulation results demonstrate
that diffusion can be enhanced, as well as, suppressed to a large extent by suitably adjusting direction and amplitude of the driving force. In the very low forcing region, active particles are more mobile than passive ones. For the self-propulsion persistent length larger than the channel compartment size,  both the diffusivity and mobility are insensitive to the forcing as long as $F<v_0$. On the other hand, with gradually increasing the strength of drive, mobility of passive particles increases and even surpasses active particles. For quite a large range of forcing, active species diffuse less and move faster through the array irrespective of the direction of the drive. These features could allow segregating active particles  from the passive ones by driving their mixture through an array of obstacles. Further, larger mobility to diffusivity ratio is desirable for many nano-technological and biomedical applications. 

For $F > v_0$, impacts of trapping, depinning and focusing action of drive are manifested through excess diffusivity peaks, abrupt change in mobility with driving strength and negative differential mobility. The depinning  mechanism largely depends on the persistence length of self-propulsion. As expected, when persistence length is very short, even smaller than the channel bottleneck, active species behave like a passive one. Impact of channel geometry has been analysed to better understand transport of active particles. Although our study simplifies the problem through the assumptions of point-like particles and circular obstructions with their regular arrangement, we still expect that the transport features which our simulation results demonstrate are quite robust and can appear in the experiments.      

\section*{Acknowledgments} P.K.G. is supported by SERB Core
Research Grant No. CRG/2021/007394. P.B. thanks UGC, New Delhi, India, for the award of a Junior Research Fellowship. 

\section*{Data Availability}
The data that support the findings of this study are available within the article.

\section*{Conflict of interest}
The authors have no conflicts to disclose.

\end{document}